\documentclass[showpacs,preprintnumbers,amsmath,amssymb]{revtex4}
\usepackage{epsf}
\usepackage[dvips]{graphicx}
\usepackage[dvips]{color}
 \newcommand{\eins}{\mbox{$1 \hspace{-1.0mm}  {\bf l}$}}
 
 \newcommand{\be}{\begin{equation}}
 \newcommand{\ee}{\end{equation}}
 \newcommand{\bea}{\begin{eqnarray}}
 \newcommand{\eea}{\end{eqnarray}}
 \newcommand{\ket}[1]{ | \, #1  \rangle}
 
 \newcommand{\bra}[1]{ \langle #1 \,  |}
 
 \newcommand{\proj}[1]{\ket{#1}\bra{#1}}

\def\kk{\rangle\!\rangle}\def\bb{\langle\!\langle}
\def\Tr{\operatorname{Tr}}

\def\>{\rangle}\def\<{\langle}
\def\sH{\mathcal{H}}
\def\sK{\mathcal{K}}
\def\sL{\mathcal{L}}
\begin{document}
\title{Economical Phase-Covariant Cloning of Qudits} \author{Francesco
  Buscemi}\email{buscemi@fisicavolta.unipv.it} \author{Giacomo Mauro
  D'Ariano}\email{dariano@unipv.it} \author{Chiara
  Macchiavello}\email{chiara@unipv.it} \affiliation{QUIT Group,
  Dipartimento di Fisica ``A.  Volta'', Universit\`a di Pavia, via
  Bassi 6, I-27100 Pavia, Italy}\homepage{http://www.qubit.it/}
\date{July 14, 2004}
\begin{abstract}
  We derive the optimal $N\to M$ phase-covariant quantum cloning for
  equatorial states in dimension $d$ with $M=kd+N$, $k$ integer. The
  cloning maps are optimal for both global and single-qudit fidelity.
  The map is achieved by an ``economical'' cloning machine, which
  works without ancilla.
\end{abstract}

\maketitle

\section{Introduction}

In quantum information the study of optimal cloning machines is a
focus of interest since, by definition, cloning is synonymous of
multiplexing quantum information, which has limitations in principle
by the no-cloning theorem \cite{no-cloning,yuenclon}. In the large
variety of proposals for optimal cloners, the fidelity of the machine
depends on the choice of input states, with the machine often working
in a covariant way, producing ``rotated'' clones from rotated inputs.
In particular, the case of $\mathbb{SU}(d)$ covariance corresponds to
universal cloning \cite{Buz-Hill,Gisin-Massar,werner}, with equal
fidelity for all unitarily connected states, e. g. for all pure
states.  Clearly, by taking smaller input set of states the cloning
performance can be improved, e. g. for smaller covariance groups. Also
in connection with the eavesdropping strategies in BB84 quantum
cryptography \cite{BB84}, the phase-covariant cloning of equatorial
states has been extensively studied for qubits \cite{bcdm,opt_pcc},
and more generally qudits \cite{fan-phcov-d}, the latter also with the
motivation of understanding which features are peculiar of dimension
two.

In this paper we will consider multi-phase covariant cloning
transformations on ``equatorial'' states
\begin{equation}
|\psi(\{\phi_j\})\>=\frac{1}{\sqrt d}(\ket{0}+e^{i\phi_1}\ket{1}
+e^{i\phi_2}\ket{2}+...+e^{i\phi_{d-1}}\ket{d-1}),\label{qudits}
\end{equation}
where the $\phi_j$'s are independent phases in the interval
$[0,2\pi)$. An issue which recently has attracted interest in the
literature is the possibility of achieving the cloning without the
need of an ancilla---a so-called ``economical'' cloning
\cite{economical}. As we will see in the following, the multi-phase
covariant cloning machines are indeed economical for $M=kd+N$ output
copies, $k$ integer.

The paper is organized as follows. In Section \ref{ph-cov-clon}, after
introducing the notations and the basic definitions taken from Refs.
\cite{opt_pcc,multi-phase}, we describe the general approach to
covariant cloning maps of Ref.  \cite{dalop}, and apply it to the case
of $N\to M$ phase-covariant cloner. In Section \ref{economical-maps}
we give a brief formalization of the concept of ``economical maps'' by
means of the Stinespring representation theorem for completely
positive maps. In Section \ref{one-to-M} we explicitly find the
optimal $1\to M$ cloner for both single-qudit fidelity and global
fidelities. In Section \ref{N-to-M} we generalize all previous results
to the case $N\to M$. Section \ref{conclusions} concludes the paper
with a comparison of fidelities in the various cases.

\section{Phase-Covariant Cloning}\label{ph-cov-clon}
We want to derive the optimal $N\to M$ cloning transformations
$\mathcal{C}$ that are covariant under the group of rotations of all
the $d-1$ independent phases $\{\phi_j\}$, $\phi_j\in[0,2\pi)$,
\begin{equation}
U(\{\phi_j\})=|0\>\<0|+\sum_{j=1}^{d-1}|j\>\<j|e^{i\phi_j},
\end{equation}
where $\{\ket{0},\ket{1},\ket{2}...\ket{d-1}\}$ represents a basis for
the $d$-dimensional Hilbert space $\sH$ of the system of a single
copy.  We will restrict the study of such maps to the set of the
$N$-fold tensor product of generalized equatorial pure states
\begin{eqnarray}
U(\{\phi_j\})|\psi_0\>\doteq|\psi(\{\phi_j\})\rangle,\label{qudit1}
\end{eqnarray}
with $|\psi(\{\phi_j\})\rangle$ given in Eq. (\ref{qudits}).
Here $\ket{\psi_0}$ is the equatorial superposition
\begin{equation}
\ket{\psi_0}=d^{-1/2}\sum_i|i\>.\label{psi0}
\end{equation}
The choice $\phi_0=0$ is not restrictive, since an overall phase is
negligible.

As argued in Ref.  \cite{opt_pcc}, we consider cloning maps for which
the $N$-copy input state and the $M$-copy output state are both
supported on the symmetric subspaces $\sH^{\otimes N}_+$ and
$\sH^{\otimes M}_+$, respectively. We choose orthonormal basis in the
symmetric subspace of the form
\begin{equation}\label{symmetrization}
|\{n_i\}\>\doteq\ket{n_0,n_1,n_2,...n_{d-1}}=\frac{1}{\sqrt{N!}}\sum_{\{\pi\}}P_\pi^{(N)}|\underbrace{00\dots0}_{n_0}\underbrace{11\dots1}_{n_1}\dots\underbrace{d-1\dots d-1}_{n_{d-1}}\>,
\end{equation}
where $P_\pi^{(N)}$ denotes the permutation operator of $N$ qubits,
$n_0$ is the number of qudits in state $\ket{0}$, $n_1$ in state
$\ket{1}$, and so on, with the constraint $\sum_{i=0}^{d-1} n_i=N$ for
the input state, and, analogously, for the output state. In the whole
paper we will consistently use letters $n$'s for input and $m$'s for
output.  The covariance condition for the cloning transformation
$\mathcal{C}$ under the group of multi-phases rotations reads
\begin{equation}\label{phase-cov-cond}
\mathcal{C}\left(U(\{\phi_j\})^{\otimes N}\,\rho^{\otimes N}\,U^\dag(\{\phi_j\})^{\otimes N}\right)=U(\{\phi_j\})^{\otimes M}\,\mathcal{C}(\rho^{\otimes N})\,U^\dag(\{\phi_j\})^{\otimes M}.
\end{equation}
As proved in Ref. \cite{dalop} the covariance condition can be
conveniently studied in terms of the positive operator on
$\sH^{\otimes M}_+\otimes\sH^{\otimes N}_+$
\begin{equation}\label{Rop}
R\doteq(\mathcal{C}\otimes\mathcal{I})(|\openone\kk\bb\openone|),
\end{equation}
where $\mathcal{I}$ is the identity map and $|\openone\kk$ is the non
normalized maximally entangled vector in $\sH^{\otimes N}_+\otimes
\sH^{\otimes N}_+$
\begin{equation}\label{max-ent}
|\openone\kk=\sum_{\{n_i\}}|\{n_i\}\>\otimes|\{n_i\}\>.
\end{equation}
The correspondence $\mathcal{C}\leftrightarrow R$ between completely
positive maps and positive operators is one-to-one, and can be
inverted as follows
\begin{equation}\label{reconstruction}
\mathcal{C}(O)=\Tr_{\sH^{\otimes N}_+}\left[\left(\openone_{\sH^{\otimes M}_+}\otimes O^T\right)\ R\right],
\end{equation}
where $O^T$ denotes the transposition of the operator $O$ with respect
to the orthonormal basis in Eq. (\ref{max-ent}). Notice that for the
state $|\psi_0\>$ of Eq. (\ref{psi0}) one has
$(|\psi_0\>\<\psi_0|^{\otimes N})^T=|\psi_0\>\<\psi_0|^{\otimes N}$
since $|\psi_0\>$, by construction, has all real coefficients with
respect to the basis in Eq.  (\ref{max-ent}). The trace-preservation
condition for $\mathcal{C}$ reads
\begin{equation}\label{trace-preservation}
\Tr_{\sH^{\otimes M}_+}[R]=\openone_{\sH^{\otimes N}_+}.
\end{equation}

Following Ref. \cite{dalop}, the covariance property
(\ref{phase-cov-cond}) rewrites as a commutation relation
\begin{equation}\label{commutation}
\left[R\,,\,U(\{\phi_j\})^{\otimes M}\otimes U^*(\{\phi_j\})^{\otimes N}\right]=0,
\end{equation}
where the complex conjugated $U^*$ of $U$ is defined as the operator
having as matrix elements the complex-conjugated matrix elements of
$U$ with respect to the same orthonormal basis in Eq. (\ref{max-ent}).
Eq. (\ref{commutation}) in turn implies by Schur Lemma a block-form
for $R$
\begin{equation}\label{block-form}
R=\bigoplus_{\{m_j\}}R_{\{m_j\}},
\end{equation}
where each set of values $\{m_j\}$ identifies a unique class of
equivalent irreducible representations of $U(\{\phi_j\})^{\otimes
  M}\otimes U^*(\{\phi_j\})^{\otimes N}$. The equivalent
representations within each class can be conveniently written as
\begin{equation}
\Big\{\ket{m_0+n_0,m_1+n_1,m_2+n_2,...m_{d-1}+n_{d-1}}\ket{n_0,n_1,n_2,...n_{d-1}}\Big\}_{\{n_i\}},
\end{equation}
with $\sum_{i=0}^{d-1} n_i=N$ and $\sum_{j=0}^{d-1} m_j=M-N$. The
multi-index $\{n_i\}$ runs over all orthonormal vectors of the basis
for $\sH^{\otimes N}_+$ used in Eq. (\ref{max-ent}). With this
notation, Eq. (\ref{block-form}) becomes
\begin{equation}
R=\sum_{\{m_j\}}\sum_{\{n'_i\},\{n''_i\}}r^{\{m_j\}}_{\{n'_i\},\{n''_i\}}|\{m_j\}+\{n'_i\}\>\<\{m_j\}+\{n''_i\}|\otimes|\{n'_i\}\>\<\{n''_i\}|.
\end{equation}

In the following, in order to evaluate the optimality of the map, we will use as figures of merit
the single-qudit fidelity 
\begin{equation}\label{ssfid}
\Tr\left[\left(|\psi_0\>\<\psi_0|\otimes\openone^{M-1}\otimes|\psi_0\>\<\psi_0|^{\otimes N}\right)\,R\right]
\end{equation}
and the global fidelity
\begin{equation}\label{gfid}
\Tr\left[\left(|\psi_0\>\<\psi_0|^{\otimes M+N}\right)\,R\right].
\end{equation}
Notice that in deriving the last two equations we used the covariance
property (\ref{qudit1}) of the input states, the reconstruction
formula (\ref{reconstruction}), the commutation property
(\ref{commutation}), and the cyclic invariance of the trace.  Since
each single contribution to the single-qudit fidelity (\ref{ssfid})
and to the global fidelity (\ref{gfid}) is positive versus the indeces
$\{n'_i\}$ and $\{n''_i\}$, as we will show in the following (see Eqs.
(\ref{sing-fid-contrib-D}), (\ref{sing-fid-contrib-OD}), and
(\ref{totalNM}), \cite{note-on-positivity}), the block $R_{\{m_j\}}$
must have positive elements $r^{\{m_j\}}_{\{n'_i\},\{n''_i\}}\ge 0$,
with the off-diagonal ones as large as possible, i.e.
$r^{\{m_j\}}_{\{n'_i\},\{n''_i\}}=
\sqrt{r^{\{m_j\}}_{\{n'_i\},\{n'_i\}}}
\sqrt{r^{\{m_j\}}_{\{n''_i\},\{n''_i\}}}$, \cite{cs-ineq}. This is
equivalent to say that the blocks constituting the operator $R$ are
actually rank-one blocks, namely
\begin{equation}
R_{\{m_j\}}\propto\proj{r_{\{m_j\}}},
\end{equation}
with
\begin{equation}
|r_{\{m_j\}}\>=\sum_{\{n_i\}}r^{\{m_j\}}_{\{n_i\}}\ket{\{m_j\}+\{n_i\}}\otimes|\{n_i\}\>,
\end{equation}
and, separately imposing condition (\ref{trace-preservation}) over
every block, \cite{remark1}, $\Tr_{\sH^{\otimes
    M}_+}[\proj{r_{\{m_j\}}}]=\openone_{\sH^{\otimes N}_+}$, we get
the final form for $R$
\begin{equation}\label{final-R}
R=\bigoplus_{\{m_j\}}p_{\{m_j\}}\proj{r_{\{m_j\}}},\qquad|r_{\{m_j\}}\>=\sum_{\{n_i\}}\ket{\{m_j\}+\{n_i\}}\otimes|\{n_i\}\>,
\end{equation}
where $p_{\{m_j\}}$ are free parameters satisfying $p_{\{m_j\}}\ge 0$
and $\sum p_{\{m_j\}}=1$ in order to preserve normalization and
positivity of $R$. This means that $R$ is a convex combination of
orthogonal rank-one blocks.

In Sections \ref{one-to-M} and \ref{N-to-M} we will explicitly
optimize the map starting from the $R$ operator in Eq.
(\ref{final-R}).

\section{Economical maps}\label{economical-maps}

Let $\mathcal{M}$ be a completely positive, trace-preserving map from
states on $\sH$ to states on $\sK$. The Stinespring
representation Theorem \cite{Stinespring} says that for every
completely positive trace-preserving map it is possible to find an
auxiliary quantum system with Hilbert space $\sL$ and an isometry $V$
from $\sH$ to $\sK\otimes\sL$, $V^\dag V=\openone_\sH$, such that
\begin{equation}\label{Stine}
\mathcal{M}(\rho)=\Tr_\sL[V\rho V^\dag].
\end{equation}
Starting from Eq. (\ref{Stine}), it is always possible to construct a
unitary interaction $U$ realizing $\mathcal{M}$ \cite{Kraus,sc}:
\begin{equation}\label{unit-real}
\mathcal{M}(\rho)=\Tr_\sL\left[U(\rho\otimes|a\>\<a|)U^\dag\right],
\end{equation}
where $|a\>$ is a fixed pure state of a \emph{second} auxiliary
quantum system, say $\sL' $, such that $\sH\otimes\sL '=
\sK \otimes\sL $. The Hilbert spaces $\sL$ and
$\sL' $ are generally different, and actually play different physical roles.

We define a trace-preserving completely positive map $\mathcal{M}$ to
be \emph{economical} if and only if it admits a unitary form $U$ as
\begin{equation}\label{econ-real}
\mathcal{M}(\rho)=U(\rho\otimes|a\>\<a|)U^\dag,
\end{equation}
namely, if and only if the map can be physically realized without
discarding resources. We can simply prove that the only maps
admitting an economical unitary implementation $U$ as in Eq.
(\ref{econ-real}) are precisely those for which
\begin{equation}\label{isometrical}
\mathcal{M}(\rho)=V\rho V^\dag
\end{equation}
for an isometry $V$, $V^\dag V=\openone$. In fact,
$U(\openone_\sH\otimes|a\>)$ is an isometry from $\sH$ to $\sK\otimes\sL$, since
$(\openone_\sH\otimes\<a|)U^\dag
U(\openone_\sH\otimes|a\>)=\openone_\sH$. On the other hand, 
from Eq. (\ref{isometrical}) via Gram-Schmidt one can extend any isometry $V$ from $\sH$ to
$\sK\otimes\sL$ to a unitary $U$ on the same output space, and write it in the form 
$V=U(\openone_\sH\otimes|a\>)$ for unit vector $|a\>\in\sL' $, with 
$\sH\otimes\sL '=\sK \otimes\sL $. For a detailed discussion about the
explicit construction procedures, see Ref. \cite{sc}.

Allowing classical resources ``for free'', a map should be defined as
economical also in the case in which it admits a
\emph{random-economical} realization as
\begin{equation}
\mathcal{M}(\rho)=\sum_ip_iU_i(\rho\otimes|a\>\<a|)U^\dag_i,
\end{equation}
where $p_i\ge 0$, $\sum_ip_i=1$. Using the same fixed ancilla state
$|a\>$ for all indeces $i$ is not a loss of generality, since in
constructing the operators $U_i$'s there is always freedom in the
choice of the vector $|a\>$. According to this more general
definition, all economical maps can always be written as a
randomization of Eq. (\ref{isometrical}):
\begin{equation}
\mathcal{M}(\rho)=\sum_ip_iV_i\rho V^\dag_i.
\end{equation}

\section{Optimal $1\to M$ Cloning}\label{one-to-M}

The fidelity of the reduced output state
$\Tr_{M-1}\left[\mathcal{C}(|\psi(\{\phi_j\})\>\<\psi(\{\phi_j\})|)\right]$
with respect to the input state
$|\psi(\{\phi_j\})\>\<\psi(\{\phi_j\})|$ is given by
\begin{equation}
\Tr\left[\left(\proj{\psi_0}\otimes\eins^{M-1}\otimes\proj{\psi_0}\right)\,
R\right].
\end{equation}
In the case of $1\to M$ cloning, the $R$ operator in Eq.
(\ref{final-R}) has the following structure:
\begin{equation}
R=\bigoplus_{\{m_j\}}p_{\{m_j\}}\proj{r_{\{m_j\}}},\qquad|r_{\{m_j\}}\>=\sum_i\ket{m_0,m_1,\dots,m_i+1,\dots}\otimes|i\>,
\end{equation}
with $\sum_j m_j=M-1$, whence the form of the summands:
\begin{equation}
\begin{split}
\Tr&\left[\left(\proj{\psi_0}\otimes\eins^{M-1}\otimes\proj{\psi_0}\right)
\ket{m_0+1,m_1,m_2,...m_{d-1}}\bra{m_0,m_1+1,m_2,...m_{d-1}}\otimes\ket{0}
\bra{1}\right]\\
&=\frac{1}{d^2}\frac{(M-1)!}{m_0!m_1!...m_{d-1}!}\sqrt{\frac{(m_0+1)!m_1!....m_{d-1}!}{M!}\frac{m_0!(m_1+1)!....m_{d-1}!}{M!}}=\frac{1}{Md^2}\sqrt{(m_0+1)(m_1+1)}.
\label{fc}
\end{split}
\end{equation}
The final contribution to the partial fidelity due to the set of
equivalent representations labeled by $\{m_j\}$ is
\begin{equation}
F_{\{m_j\}}=\Tr\left[\left(\proj{\psi_0}\otimes\eins^{M-1}\otimes\proj{\psi_0}\right)\,R_{\{m_j\}}\right]=\frac{p_{\{m_j\}}}{d^2}\left[d + \frac{1}{M}\sum_{i\neq k}\sqrt{(m_i+1)(m_k+1)}\right].
\label{fn}
\end{equation}

The projector $\proj{r_{\{m_j\}}}$ that contributes most to the
fidelity is the one that maximizes the quantity $\sum_{i\neq k}
\sqrt{(m_i+1)(m_k+1)}$, with the constraint $\sum_{i=0}^{d-1}
m_i=M-1$. In the case
\begin{equation}\label{opt1}
M=dk+1,\qquad k\in\mathbb{N},
\end{equation}
the optimization gives the simple result $m_i=k$ for all $i$. The
$1\to(kd+1)$ optimal phase-covariant cloning machine is then
completely described by the rank-one positive operator
$R=|r_{\{k\}}\>\<r_{\{k\}}|$. The Kraus form of the optimal map is
then reconstructed as
\begin{equation}\label{kraus-1}
  \mathcal{C}(\rho)=\Tr_{\sH}\left[\left(\openone_{\sH^{\otimes M}_+}\otimes\rho^T\right)\,R\right]=V\rho V^\dag,
\end{equation}
where $V:\sH\to\sH^{\otimes M}_+$ is the isometry, i.e. $V^\dag
V=\openone$, defined as follows:
\begin{equation}\label{isometry-1}
V|i\>=|m_0=k,m_1=k,\dots,m_i=k+1,\dots\>.
\end{equation}

The fact that the Kraus operator describing the map is isometrical---a
consequence of $R$ being rank-one---automatically guarantees that no
additional \emph{ancillae} (other than the $M-1$ blank states) are
needed in order to unitarily realize the cloning transformation
$\mathcal{C}$.

From Eq. (\ref{fn}) one obtains the single-qudit fidelity of our
multi-phase-covariant economical cloning machine from one to $M=kd+1$
copies
\begin{eqnarray}
F_\mathcal{C}(1,M=kd+1)=\frac{1}{d}+\frac{(d-1)(M+d-1)}{Md^2}.
\label{F}
\end{eqnarray}
Notice that the above result, in the limit $M\to\infty$, is consistent
with the fidelity of optimal phase estimation on a single qudit as
worked out in Ref. \cite{multi-phase}.

An important remark that remains to be stressed is that the value of
$\{m_j\}$ maximizing the single-qudit fidelity (\ref{fn}) maximizes the
total fidelity as well. In fact, the total fidelity of the
$\{m_j\}$-th block is given by
\begin{equation}
\begin{split}
  p_{\{m_j\}}\sum_{ij}&\Tr\left[|\psi_0\>\<\psi_0|^{\otimes(M+1)}\;|\dots,m_i+1,\dots\>\<\dots,m_j+1,\dots|\otimes|i\>\<j|\right]=p_{\{m_j\}}\sum_i\left|\<\psi_0^{\otimes(M+1)}|\dots,m_i+1,\dots\>|i\>\right|^2\\ 
&=\frac{p_{\{m_j\}}}{d^{M+1}}\sum_i\frac{M!}{m_0!\dots(m_i+1)!\dots}
=\frac{p_{\{m_j\}}}{d^{M+1}}\sum_i\binom{M}{m_0;\cdots;m_i+1;\cdots}
\end{split}
\end{equation}
where, in the last line, we used the standard notation for multinomial
coefficients with the implicit constraints $\sum_j m_j=M-1$.  In order
to maximize the value of the multinomial coefficient, the vector
$\{m_j\}$ has to be as ``flat'' as possible, namely, with all entries
as close as possible to each other. The situation in which the
solution is unique and given by $m_i=k$ for all $i$ is the same as in
Eq.  (\ref{opt1}). This means that the single-qudit fidelity
optimization procedure provides the same result as the total fidelity
optimization, and the map written in Eqs.  (\ref{kraus-1}) and
(\ref{isometry-1}) is optimal in both approaches.

\section{Optimal $N\to M$ cloning}\label{N-to-M}

In the general case of arbitrary values for $N$ and $M$, the
single-qudit fidelity is obtained by summing up contributions of the
form
\begin{equation}\label{contrib}
\Tr\left[|\psi_0\>\<\psi_0|\otimes\openone^{\otimes(M-1)}\otimes|\psi_0\>\<\psi_0|^{\otimes N}\ |\{m_j\}+\{n'_j\}\>\<\{m_j\}+\{n''_j\}|\otimes|\{n'_j\}\>\<\{n''_j\}|\right]
\end{equation}
because of the block-form (\ref{final-R}) of the $R$ operator.  Before
getting into the explicit calculation for the partial fidelity, it is
possible to somehow simplify the problem by noticing that the presence
of $|\psi_0\>\<\psi_0|\otimes\openone^{\otimes(M-1)}$ in Eq.
(\ref{contrib}) restricts the evaluation of the fidelity to those
blocks for which the $M$-particles states differ \emph{at most} for a
single-particle state.

The diagonal contributions to $F_{\{m_j\}}$ (as before, the
single-qudit fidelity calculated only for the $\{m_j\}$-th block) are
then proportional to (apart from the probability $p_{\{m_j\}}$)
\begin{equation}\label{sing-fid-contrib-D}
\begin{split}
\Tr&\left[|\psi_0\>\<\psi_0|\otimes\openone^{\otimes(M-1)}\otimes|\psi_0\>\<\psi_0|^{\otimes N}\ |\dots,m_i+n_i+1,\dots\>\<\dots,m_i+n_i+1,\dots|\otimes|\dots,n_i+1,\dots\>\<\dots,n_i+1,\dots|\right]\\
&=\frac{1}{d^N}\;\frac{N!}{n_0!\dots(n_i+1)!\dots}\;\frac{1}{d}\;\frac{(m_0+n_0)!\dots(m_i+n_i+1)!\dots}{M!}\;\frac{M!}{(m_0+n_0)!\dots(m_i+n_i+1)!\dots}\\
&=\frac{1}{d^{N+1}}\;\frac{N!}{n_0!\dots(n_i+1)!\dots},\\
\end{split}
\end{equation}
where for sake of simplicity in the last equation the notation was
slightly modified with $\sum_i n_i=N-1$ while, again, $\sum_j
m_j=M-N$. The off-diagonal terms are
\begin{equation}\label{sing-fid-contrib-OD}
\begin{split}
  \Tr&\left[|\psi_0\>\<\psi_0|\otimes\openone^{\otimes(M-1)}\otimes|\psi_0\>\<\psi_0|^{\otimes N}\ |\dots,m_i+n_i+1,\dots\>\<\dots,m_j+n_j+1,\dots|\otimes|\dots,n_i+1,\dots\>\<\dots,n_j+1,\dots|\right]\\
  &=\Tr\left[|\psi_0\>\<\psi_0|\otimes\openone^{\otimes(M-1)}\ 
    |\dots,m_i+n_i+1,\dots\>\<\dots,m_j+n_j+1,\dots|\right]\\
  &\phantom{000000000000000}\times\frac{1}{d^N}\sqrt{\frac{n_0!\dots(n_i+1)!\dots}{N!}}\sqrt{\frac{n_0!\dots(n_j+1)!\dots}{N!}}\frac{N!}{n_0!\dots(n_i+1)!\dots}\;\frac{N!}{n_0!\dots(n_j+1)!\dots}\\
  &=\frac{1}{d^{N+1}}\;\frac{N!}{n_0!\dots n_i!\dots n_j!\dots}\sqrt{\frac{1}{(n_i+1)(n_j+1)}}\sqrt{\frac{(m_0+n_0)!\dots(m_i+n_i+1)!\dots}{M!}}\\
  &\phantom{000000000000000}\times\sqrt{\frac{(m_0+n_0)!\dots(m_j+n_j+1)!\dots}{M!}}\frac{(M-1)!}{(m_0+n_0)!\dots(m_i+n_i)!\dots(m_j+n_j)!\dots}\\
  &=\frac{1}{Md^{N+1}}\;\frac{N!}{n_0!\dots n_i!\dots
    n_j!\dots}\sqrt{\frac{(m_i+n_i+1)(m_j+n_j+1)}{(n_i+1)(n_j+1)}}.
\end{split}
\end{equation}
At the end, the single-qudit fidelity is the sum of contributions of
the form
\begin{equation}\label{partialNM}
F_{\{m_j\}}=\frac{p_{\{m_j\}}}{d^{N+1}}\sum_{\{n_j\}}\left[\sum_i\frac{N!}{n_0!\dots(n_i+1)!\dots}+\frac{1}{M}\sum_{i\neq j}\frac{N!}{n_0!\dots n_i!\dots n_j!\dots}\sqrt{\frac{(m_i+n_i+1)(m_j+n_j+1)}{(n_i+1)(n_j+1)}}\right].
\end{equation}
As done before for the $1\to M$ cloning, in order to find the block
of $R$ realizing the optimal map, we have to maximize the off-diagonal
quantity
\begin{equation}\label{to-maximize}
\sum_{i\neq j}\sqrt{\frac{(m_i+n_i+1)(m_j+n_j+1)}{(n_i+1)(n_j+1)}}
\end{equation}
with the constraints $\sum_i n_i=N-1$ and $\sum_j m_j=M-N$. The
maximization of fidelity corresponds to maximize the quantity in Eq.
(\ref{to-maximize}) versus the variables $m_i$'s. Since the variables
$n_i$'s are summed up in Eq. (\ref{partialNM}), then the fidelity is
invariant under their permutation. Therefore the evaluation of the
maximum of the quantity (\ref{to-maximize}) resorts to maximize it for
equal $n_i$'s, whence also all $m_i$'s will be equal,
$m_i=m_*\doteq(M-N)/d$, $\forall i$. Generally, in this way one
obtains a non integer value of $m_*$, while the maximum for integer
$m_i$'s is very degenerate, since the maximum will be obtained for
unequal $m_i$'s in place of a common fractional value. This leads to
many blocks for $R$ contributing in the same way to the optimal map,
which makes the evaluation very complicate. On the other hand, the
evaluation simplifies greatly when the maximum is achieved for integer
$m_*=k$, and this corresponds to the following relation between $N$
and $M$
\begin{equation}\label{optblock}
M=kd+N.
\end{equation}

Hence, the optimal phase-covariant $N\to (N+kd)$ cloning map is
described by the rank-one operator
\begin{equation}\label{optmap1}
R=|r_{\{k\}}\>\<r_{\{k\}}|,
\end{equation}
where
\begin{equation}\label{optmap2}
|r_{\{k\}}\>=\sum_{\{n_j\}}|k+n_0,\dots,k+n_i,\dots\>_M
|n_0,\dots,n_i\dots\>_N,\qquad\sum_jn_j=N,
\end{equation}
and its single-qudit fidelity is given by
\begin{equation}\label{FNM}
F_\mathcal{C}(N,M=kd+N)=\frac{1}{d}+\frac{1}{Md^{N+1}}\sum_{\{n_j\}}\sum_{i\neq j}\frac{N!}{n_0!\dots n_i!\dots n_j!\dots}\sqrt{\frac{(n_i+k+1)(n_j+k+1)}{(n_i+1)(n_j+1)}},\qquad\sum_jn_j=N-1.
\end{equation}

Notice that being $R$ rank-one, the optimal map
derived here is again described by only one isometric Kraus operator
$V:\sH^{\otimes N}_+\rightarrow\sH^{\otimes M}_+$
\begin{equation}\label{clon-map2}
\mathcal{C}(\rho^{\otimes N})=\Tr_{\sH^{\otimes N}_+}\left[\left(\openone_{\sH^{\otimes M}_+}\otimes(\rho^{\otimes N})^T\right)\ |r_{\{k\}}\>\<r_{\{k\}}|\right]
=V\rho^{\otimes N}V^\dag,\qquad V^\dag V=\openone^{\otimes N},
\end{equation}
where the isometry $V$ acts as follows:
\begin{equation}\label{clon-map3}
V|n_0,n_1,\dots,n_i,\dots\>_N=|n_0+k,n_1+k,\dots,n_i+k,\dots\>_M.
\end{equation}
Similarly to the case $1\to (kd+1)$, the fact that the optimal $N\to
(N+kd)$ cloning map is isometrical implies that no additional ancilla is needed to unitarily realize
the map other than the $M-N$ blank copies, and (\ref{clon-map3}) is again an economical cloning
machine. 

As in Section \ref{one-to-M}, it is possible to prove that the value
of $\{m_j\}$ maximizing the single-qudit fidelity maximizes the total
fidelity as well
\begin{equation}\label{totalNM}
\begin{split}
&\sum_{\{m_j\}}p_{\{m_j\}}\sum_{\{n'_j\},\{n''_j\}}\Tr\left[|\psi_0\>\<\psi_0|^{\otimes(M+N)}\;|\{m_j\}+\{n'_j\}\>\<\{m_j\}+\{n''_j\}|\otimes|\{n'_j\}\>\<\{n''_j\}|\right]\\
=&\sum_{\{m_j\}}p_{\{m_j\}}\sum_{\{n_j\}}\left|\<\psi_0^{\otimes(M+N)}|m_0+n_0,m_1+n_1,\dots\>|n_0,n_1,\dots\>\right|^2\\
=&\frac{1}{d^{M+N}}\sum_{\{m_j\}}p_{\{m_j\}}\sum_{\{n_j\}}\binom{M}{m_0+n_0;m_1+n_1;\cdots}\;\binom{N}{n_0;n_1;\cdots},
\end{split}
\end{equation}
with the usual constraints $\sum_i n_i=N$ and $\sum_j m_j=M-N$
implicit in the multinomial notation. Following the argument of the
previous Section, it is clear that the map in Eqs. (\ref{clon-map2})
and (\ref{clon-map3}) maximizing the single-qudit fidelity
(\ref{partialNM}), also maximizes the global fidelity (\ref{totalNM}).

As already noticed in the previous Section for $N=1$, the fidelity
(\ref{FNM}), in the limit of an infinite number of output copies,
namely $k\to\infty$, takes the form (in the limit, $M\approx kd$)
\begin{equation}\label{FPEN}
F_\textrm{pe}(N)=\frac{1}{d}+\frac{1}{d^{N+2}}\sum_{\{n_i\}}\sum_{i\neq j}\frac{N!}{n_0!\dots}\frac{1}{\sqrt{(n_i+1)(n_j+1)}}.
\end{equation}
The above expression coincides with the fidelity of optimal
multi-phase estimation on equatorial qudits derived in Ref.
\cite{multi-phase}.

\section{Conclusions}\label{conclusions}
\begin{figure}
\begin{center}
\includegraphics{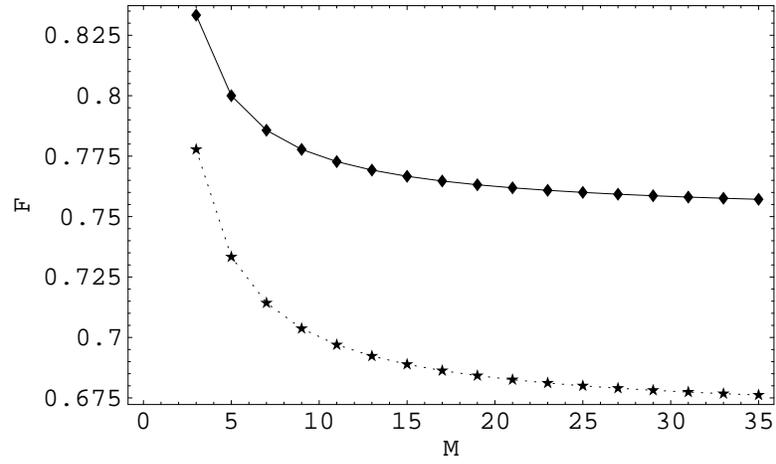}
\caption{Single-qubit fidelity for $1\to M$ cloning:  phase-covariant (continuous line); universal
  (dotted line). The numerical results are equal to those in Refs.
  \cite{werner} and \cite{opt_pcc}.}
\label{fig1} 
\end{center}
\end{figure}

\begin{figure}
\begin{center}
\includegraphics{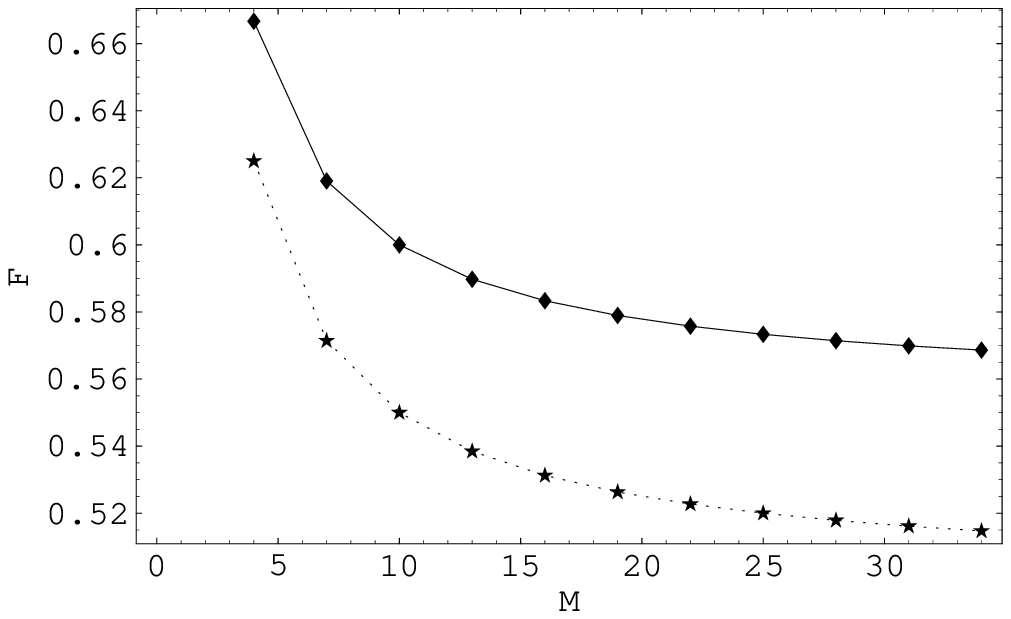}
\caption{Single-qutrit fidelity for $1\to M$ cloning:  multi-phase (continuous line); universal 
(dotted line).} \label{fig2}
\end{center}
\end{figure}

\begin{figure}
\begin{center}
\includegraphics{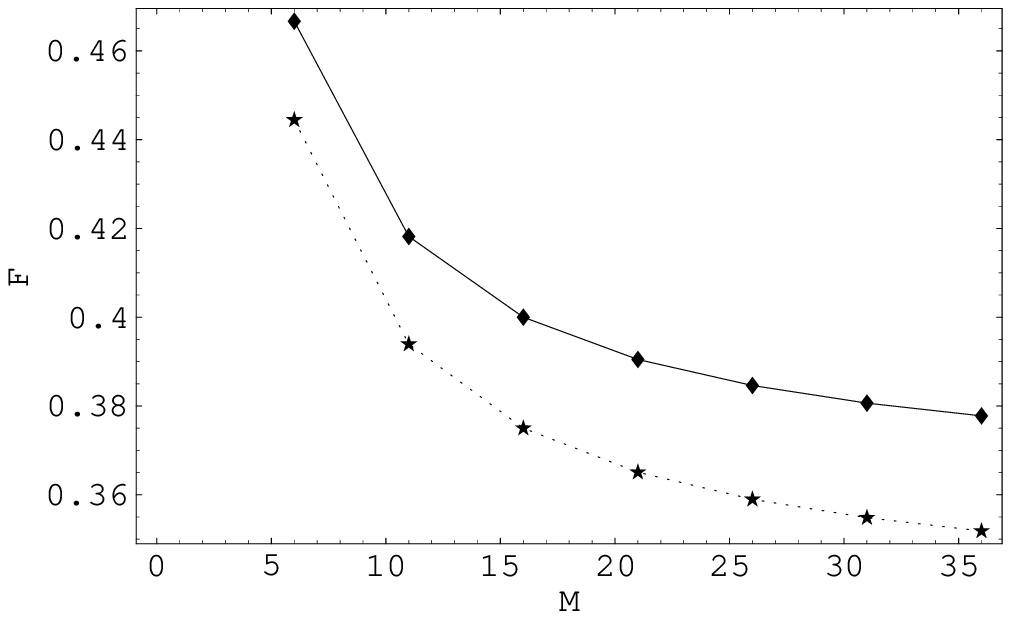}
\caption{Single-qudit, $d=5$, fidelity for $1\to M$ cloning:  multi-phase (continuous line);
  universal (dotted line).} \label{fig3}
\end{center}
\end{figure}

\begin{figure}
\begin{center}
  \includegraphics{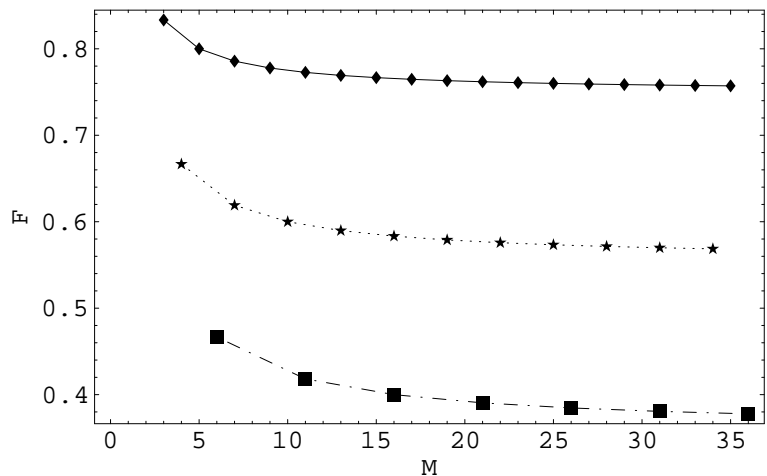}
\caption{Comparison between single-qudit fidelities for multi-phase covariant $1\to M$ cloning:
  $d=2$ (continuous line); $d=3$ (dotted line); $d=5$ (dash dotted
  line). The three curves for large $M$ approach the fidelity of
  optimal phase estimation over one copy, namely $F=3/4$ for qubits,
  $F=5/9$ for qutrits, and $F=9/25$ for $d=5$.}
\label{fig4}
\end{center}
\end{figure}

\begin{figure}
\begin{center}
  \includegraphics{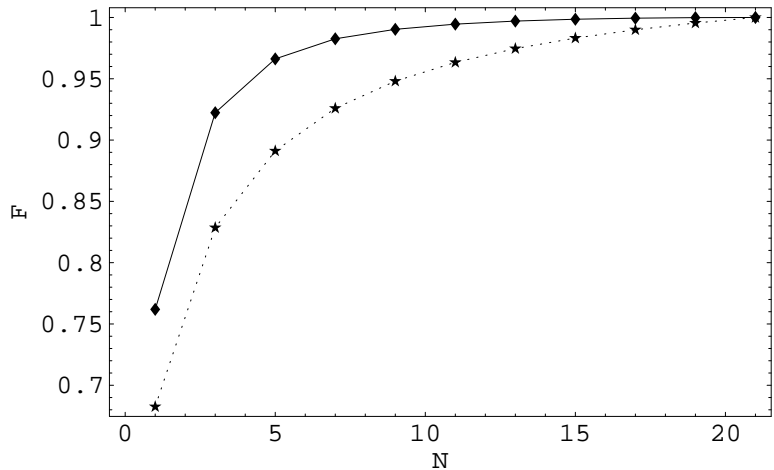}
\caption{Saturation of single-qubit fidelity for phase-covariant (continuous line) and universal (dotted
  line) $N\to M$ cloning machines versus $N$, for $M=21$.} \label{fig5}
\end{center}
\end{figure}

\begin{figure}
\begin{center}
\includegraphics{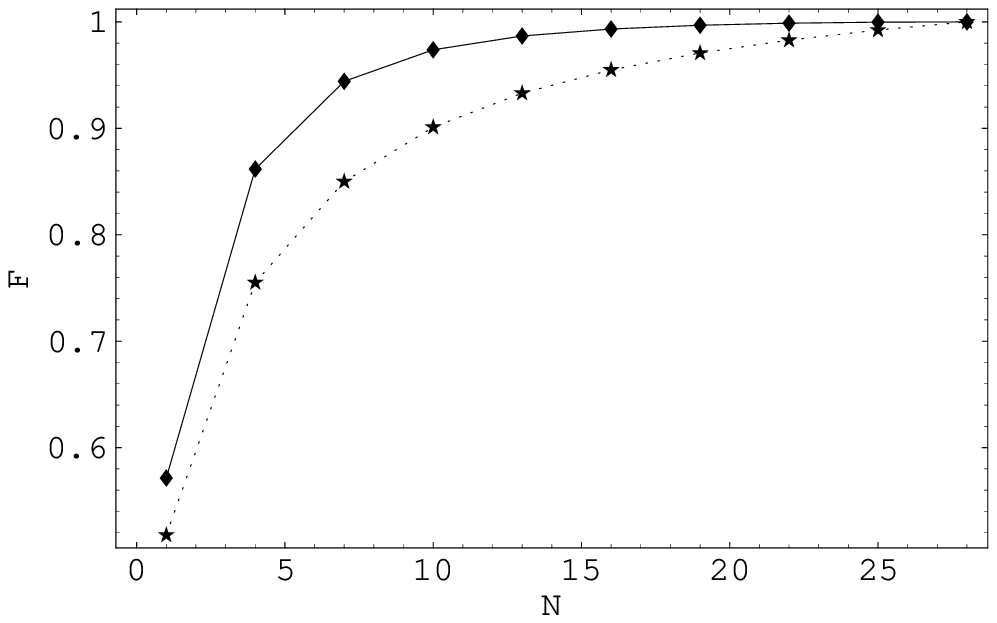}
\caption{Saturation of single-qutrit fidelity for multi-phase (continuous line) and universal
  (dotted line) $N\to M$ cloning machines versus $N$, for $M=28$.} \label{fig6}
\end{center}
\end{figure}

\begin{figure}
\begin{center}
\includegraphics{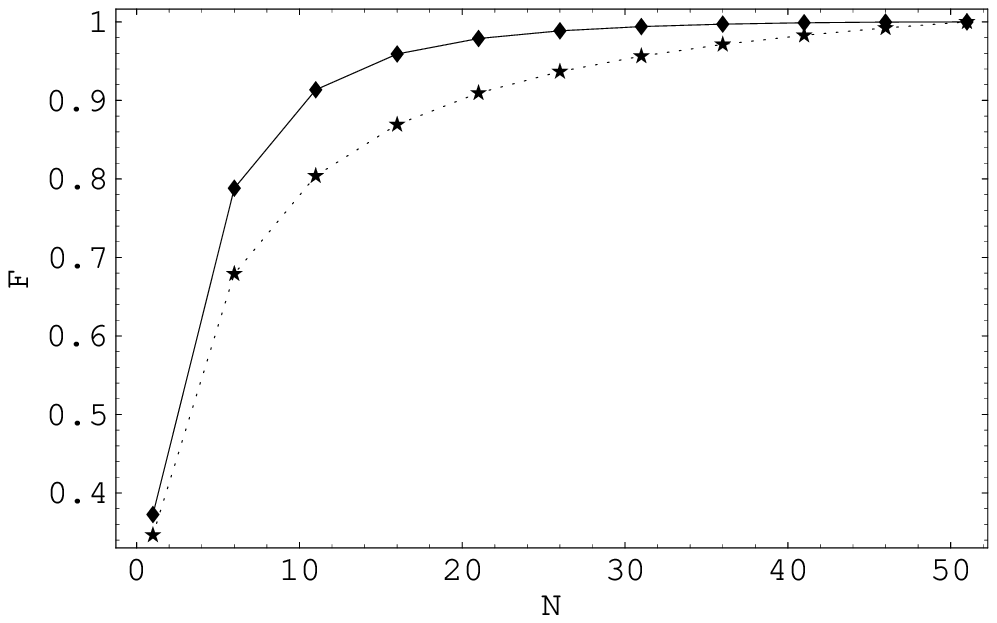}
\caption{Saturation of single-qudit, $d=5$, fidelity for multi-phase (continuous line) and universal
 (dotted line) $N\to M$ cloning machines versus $N$, for $M=51$.} \label{fig7} 
\end{center}
\end{figure}

We have addressed the problem of optimal phase-covariant cloning with
multiple phases for qudits, with arbitrary number of input copies $N$
and output copies $M$. The optimization greatly simplifies for values
of $M$ and $N$ related as $M=kd+N$, with $k$ integer. The cloning maps
are optimal for both global and single-qudit fidelity. The map is
achieved by an economical cloning machine, which works without
ancilla. We have evaluated the asymptotic behaviour of the fidelity
for large $M$, and recovered the fidelity of optimal multi-phase
estimation \cite{multi-phase}. In Figs.  \ref{fig1}, \ref{fig2}, and
\ref{fig3} it is possible to compare the single-qudit fidelities of
multi-phase covariant and universal covariant $1\to M$ cloning
machines.  Increasing $M$ the fidelities tend to the limit of optimal
phase estimation and state estimation fidelity, respectively.
Increasing the dimension $d$ of the Hilbert space, the quality of the
clones gets worse, as plotted in Fig. \ref{fig4}. Actually, for fixed
$N$ and $M$, the single-qudit fidelity of the cloner goes to zero with
$d^{-1}$, as turns out from Eq.  (\ref{FNM}). On the other hand, for
fixed $M$, the fidelity saturates to one as $N$ gets close to $M$, in
both multi-phase and universal covariant frameworks, as one can see in
Figs. \ref{fig5}, \ref{fig6}, and \ref{fig7}. As general remark,
notice that, increasing the dimension of the input system for fixed
$N$ and $M$, the fidelities of multi-phase and universal cloners
become closer to each other.


\appendix

\section*{Acknowledgments}

This work has been jointly funded by the EC under the programs ATESIT
(Contract No. IST-2000-29681), SECOQC (Contract No. IST-2003-506813)
and INFM PRA-CLON.


\begin{thebibliography}{}
\bibitem{note-on-positivity} More precisely, the single-qudit fidelity
  (\ref{ssfid}) and the global fidelity (\ref{gfid}) are linear
  functionals in $r^{\{m_j\}}_{\{n'_i\},\{n''_i\}}$ of the form
  $\sum_{\{m_j\}} \sum_{\{n'_i\},\{n''_i\}}
  a^{\{m_j\}}_{\{n'_i\},\{n''_i\}} r^{\{m_j\}}_{\{n'_i\},\{n''_i\}}$
  with positive $a^{\{m_j\}}_{\{n'_i\},\{n''_i\}}$.
\end{thebibliography}
\end{document}